# Science on YouTube: What users find when they search for climate science and climate manipulation

Joachim Allgaier
Institute of Science, Technology and
Society Studies
Alpen-Adria-Universität Klagenfurt
Sterneckstraße 15
9010 Klagenfurt, Austria
Phone: +43/463/2700-6157
E-Mail: joachim.allgaier@aau.at

## ABSTRACT
Online video-sharing sites such as YouTube are very popular and also used by a lot of people to obtain knowledge and information, also on science, health and technology. Technically they could be valuable tools for the public communication of science and technology, but the users of YouTube are also confronted with conspiracy theories and erroneous and misleading information that deviates from scientific consensus views. This contribution details the results of a study that investigates what kind of information users find when they are searching for climate science and climate manipulation topics on YouTube and whether this information corresponds with or challenges scientific consensus views. An innovative methodological approach using the anonymization network *Tor* is introduced for drawing randomized samples of YouTube videos. This approach was used to select and examine a sample of 140 YouTube videos on climate topics.

## 1. INTRODUCTION
The online video-sharing website YouTube has been a phenomenal success and growing rapidly since its launch in 2005. YouTube today is one of the most popular internet sites and also the second most popular search engine used after Google in many countries [19]. According to the self-description of YouTube it has over a billion users, almost one-third of all people on the Internet [21]. The research presented in this contribution is particularly interested in the role of online video-sharing platforms, such as YouTube, for the public communication of science. Many citizens do use YouTube as a source of information about issues concerning science, technology and medicine [2]. Research has shown that high reading levels are required to comprehend web-based textual information on science, technology and medicine, and that might be a reason why many people prefer to use and watch YouTube videos in order to find information about scientific and other issues that interest them [6]. YouTube is particularly popular among young people, for instance a recent study in Germany found that 94 percent of youths between 12 and 19 years are on YouTube and that 81 percent use it regularly [13]. Another study from Germany [9] found that more than two thirds (69 percent) of questioned young people between 14 and 29 years said they use YouTube (and other online video platforms) to get informed about science and research. Among those between 30 and 39 years more than half (55 percent) said the same, and among those between 40 and 49 years it is still almost half (46 percent) who get informed via YouTube. When YouTube is so influential and so many people are using YouTube to get informed about science, technology and medicine, the big question is what kind of information do they find there and how the quality of information of YouTube is. The video format has a great potential for disseminating knowledge, it allows using visual and audio channels in isolation or combined for transmitting text, images, animations, films, subtitles, multiple languages and many other innovative and creative means of communication. Technically it could be a powerful tool for education and science, technology and health communication. However, various studies, mainly from the area of health communication have shown, that the quality of information on YouTube about biomedical topics strongly varies and that it is often strongly biased and, from a biomedical or scientific point of view, often inaccurate or erroneous [1]. For instance, one recent study compared information about a reported link between MMR vaccination and the development of autism in children, for which there is no scientific evidence, on YouTube, Google, Wikipedia and the scientific database PubMed [18]. The study authors found that from a biomedical point of view the lowest quality of information was found on YouTube and that the incorrect information also stayed there without being corrected for the longest time. YouTube is also notorious as a sort of Eldorado for conspiracy theories and other highly controversial content, for instance about the Ebola Virus disease [4]. One reason for this is that YouTube is a social media site without any quality or editorial control; virtually everybody can open an account and upload content on this platform [17]. Another reason is that video formats have become extremely popular and rapid technological advances and mobile technologies allow more and more people not just to watch videos, but also to produce them themselves. (Science) communication and webscience research has just begun to study YouTube and other online video-sharing websites empirically and there is still a massive gap in the research literature on what kind of content users find there, how they find it there, who uploads content with what kinds of intentions and how different groups of people perceive and make sense of the content they find on online video-sharing websites, such as YouTube. This contribution wants to address this gap by presenting results from a study about science content on YouTube. A novel methodological approach will be introduced involving the anonymity network *Tor*, which has been applied to find out more about what kind of information concerning climate change, climate science and climate manipulation can be found on YouTube, which are topics that are highly controversial in the eyes of many people. The central aim of this research is to find out if this information found on YouTube corresponds with mainstream scientific positions or if it challenges scientific consensus views.

## 2. METHOD

Studying YouTube empirically is still challenging. There are many elements that could be studied, for instance the genre of videos, the user statistics, the recommender system of YouTube, or the comments from other users [11]. In health communication research it has become something like a convention to take, for instance, the 100 videos that have received most views and that appear after entering a particular search term or search string and to analyze these videos for their medical accuracy [5]. This approach, however, is not going to provide answers to the question what people will find, if they use YouTube as a search engine, but only to the question what the quality of videos is that have been viewed most often, no matter how users found them (e.g. through recommendations on websites, links on twitter, email recommendation from friends etc.). Sampling online content via search engines is difficult, since personalized searches and the filter bubble or echo chamber [15] problem will most likely distort the results and lead to various biases [8]. An innovative methodological approach presented in this contribution is to use the anonymity network *Tor* (The Onion Router, https://www.torproject.org/, 02/02/2016) in order to alleviate potential biases that are created by single personalized searches. The free software and open network *Tor* directs Internet traffic through a free, worldwide, volunteer network of thousands of relays. Location, destination and IP address are encrypted multiple times in this process through randomly selected *Tor* relays and only the final relay decrypts the innermost layer of encryption and sends the original data to its destination without revealing the source IP address [20]. In the research presented here the *Tor* software has been installed and used to search for various keywords on YouTube relating to climate science, climate change, and climate manipulation. Each search has been repeated three times using default search settings in English, each time with a new identity provided by the *Tor* anonymity network. This procedure has been applied in order to obtain a randomized sample and to circumvent the filter bubble problem. All search results have been recorded. If more than eighty percent of the results in the third search were the same as in the first and second search (which was the case in all searches conducted), the results of the third search from YouTube for a particular search term were taken as the basis for the analysis of the first twenty videos that YouTube provided on the result page. This research strategy was adopted in order to have an approximation on what an average internet user will find if she or he searches for a particular term on YouTube. The first twenty results encompass all the results that a YouTube search provides for a particular search term on its first result page. These twenty videos have been selected and analyzed because they are most likely to be viewed by the users. Seven search terms were used to find videos on YouTube. These were 1. Climate, 2. Climate Change, 3. Climate Science, 4. Climate Engineering, 5. Geoengineering, 6. Climate Hacking, 7. Chemtrails. The last two terms are not scientific terms, but terms that are often used by opponents of mainstream science. They have been included in the sample to find out whether these searches lead to fundamentally different results than the previous terms and whether they support or challenge mainstream science. The first twenty videos have then been selected for each search, so that the sample of this study consists of 140 videos, twenty videos for each search term. The searches have been carried out between January 31, 2015 and June 25, 2015. Each video has been viewed at least once and a little summary of each video has been written down and archived, together with all textual information that YouTube provides for each video. The summary of each video has then been categorized in three different categories of videos: 1.) Videos supporting mainstream science and the scientific consensus view on human induced climate change as detailed by the Intergovernmental Panel on Climate Change (IPCC); 2.) Discussion and debate formats where mainstream science is discussed with opponents; 3.) Videos showing denial of scientific mainstream positions, such as denial of human-induced climate change or straightforward conspiracy theories about science and technology.

## 3. RESULTS

140 videos have been analyzed and categorized in three different categories. The oldest video in the sample was uploaded on September 22, 2008 and is titled: "Basics of Geography: Climate", which has received 274,071 view until January 31, 2015. This video can be classified as a science education video, which clearly supports a scientific mainstream position. The most recent video included in the sample was uploaded on June 17, 2015 and is titled: "Rush: Pope's stance on Climate Science proves He's a Marxist". This video is critical to the scientific consensus on human-induced climate change and has been viewed 13,569 until June 25, 2015. The videos included in the sample encompass different types of styles and genres, but most of them are either snippets or excerpts from previously broadcast TV programs or self-made videos. A few of the videos of the sample are also public talks and academic presentations. Table 1 shows the distribution of the videos over the three categories that have been created to categorize the videos along their stance towards mainstream scientific positions.

**Table 1. Distribution of the Videos of the Sample (n = 140)**

| Search Term | Scientific Consensus | Debate | Conspiracy or Climate Change Denial | Sum |
|---|---|---|---|---|
| *Climate* | 18 | | 2 | 20 |
| *Climate Change* | 18 | 1 | 1 | 20 |
| *Climate Science* | 17 | 1 | 2 | 20 |
| *Climate Engineering* | 8 | | 12 | 20 |
| *Geoengineering* | 2 | | 18 | 20 |
| *Climate Hacking* | 6 | 1 | 13 | 20 |
| *Chemtrails* | 1 | | 19 | 20 |
| **Sum** | 70 | 3 | 67 | 140 |

For the first three search terms Climate, Climate Change and Climate Science, the absolute majority of videos in the sample adhere to the scientific consensus view. Most of them are parts from news programs or documentaries that underline the serious consequences of man-made climate change, and many of the clips feature quotes or comments from eminent climate scientists. Many of these videos could be very helpful tools and very valuable contributions for science education and public communication and discussion of issues such as climate change. Very few videos (5 of 60) challenge mainstream scientific positions, and even less videos (2 of 60) are discussion formats in which climate scientists discuss climate change with climate change deniers. However, the picture changes entirely if we focus on the videos that appear as results in the search for Climate Engineering, Geoengineering, Climate Hacking and Chemtrails. Here more than half of the videos (62 of 80) oppose scientific consensus views or promulgate straightforward non-scientific conspiracy theories. Here we mostly find self-made and amateur videos of protagonists that believe in the so-called chemtrail conspiracy theory, which claims that evil forces spray the population with toxic substances from airplanes, but also a range of videos from people who deny man-made climate change for various reasons. In the chemtrail case the same protagonists appear over and over again in many of the videos. The name of the conspiracy is derived from the condensation trails of airplanes in the sky, which conspiracy theorists take as evidence for clandestine government or science operations against the population and they call them chemtrails. Researcher Rose Cairns from the University of Sussex studied the chemtrail movement and classifies its worldview as that of a world conspiracy theory that includes the belief in a powerful, evil and clandestine group that aspires to global hegemony [7], a position that is clearly far off scientific mainstream positions. Here the results for the search term Geoengineering are particularly striking in this this context: 90 percent of the search results adhere to the chemtrail conspiracy theory. However, both the terms Climate Engineering and Geoengineering stem from scientists and scientific discussion about how to deal with or mitigate the consequences of anthropogenic climate change with technical means (without any reference to the so-called chemtrails) [16]. Compared to other scientific fields it is a rather young epistemic community that seriously considers using technoscientific means to deal with the consequences of climate change [14]. The search term Climate Hacking addresses various issues, it is a non-scientific term for climate manipulations but also refers to hacked emails from climate scientists, that climate change deniers (unsuccessfully) used as evidence against human-induced climate change. Most of the videos that show up as results here are clearly challenging mainstream scientific positions. Another interesting result is that 95 percent of the videos that came up as results for the search term chemtrails came from users who believe in the chemtrail conspiracy and there are virtually no attempts to challenge the conspiracy theory in any way in the sample. In sum, it seems fair to say that the specific search term users make use of in YouTube searches matters and will determine to what degree they will be exposed to mainstream scientific positions or not.

## 4. DISCUSSION

The results of this research show that in the case of climate science and climate manipulation rather general search terms such as Climate, Climate Change, or Climate Science are likely to bring up videos as results that confront the users in their majority with mainstream scientific positions on human-induced climate change. These results indicate that YouTube could be a very valuable tool for informing citizens about science for some key issues. However, more specific search terms, such as Climate Engineering, Geoengineering, Climate Hacking, or Chemtrails largely led to videos that confront the users with positions that challenge mainstream scientific positions on climate change, or to outspoken conspiracy theories about science and technology – an issue which poses a major challenge to the public communication of science and technology. The later is particularly the case if users search for Geoengineering on YouTube. The chemtrail conspiracy theorists very successfully occupied this term and re-labeled their conspiracy worldview using a relatively new scientific term, making their concern sound more scientific and possibly more reasonable. This strategy also has the advantage that chemtrailers can now jump on the bandwagon when there are actual scientific discussions and events addressing technical options of climate manipulation. In fact chemtrailers explicitly address their followers to use the more scientific terms, in order not to be immediately identified as conspiracy theorists [7], as on one of their websites [10]: "The geoengineering term is related to hard science, the "chemtrails" term has no such verifiable basis but rather leads anyone that Googles the term straight to "conspiracy theory" and "hoax" definitions. Use the terms "climate engineering" and "geoengineering"." Social media websites and video platforms without editorial control, such as YouTube, provide a very fertile ground for conspiracy theorists and opponents of mainstream science because there are no gatekeepers and hence no quality control is taking place on this channels. It has been shown previously that other groups that oppose mainstream science – such as creationist groups that oppose the theory of evolution for religious reasons – call their followers to make use of YouTube as an effective tool for "internet evangelism" [3]. In this context it should also be mentioned that videos from chemtrailers, creationists and other opponents of mainstream science are often "mirrored" by their followers, this means that whole videos or parts thereof are also uploaded by various followers and friends, often under various names and with different tags and keywords, so that it is virtually impossible to dam up or delete the content once it has been uploaded. This practice is also applied in order to distort search results in favour of their own content. One striking result of this research is that no videos have been found in the sample that counter the chemtrail conspiracy theory. The part of the scientific community that seriously engages with work on climate engineering or geoengineering is so far only very marginally present on YouTube and so far it seem that the discourse on the two terms on YouTube is dominated by chemtrail conspiracy theorists. In contrast it seems that so far the mainstream scientific establishment was more successful to dominate the discourse on man-made climate change. However, these results come with caveats. The research strategy adopted here only allows for a snapshot picture on what is happening on YouTube given these terms. YouTube is a particularly lively website with heavy internet traffic that is constantly in flux and possible results may change very quickly. However, it still seems that YouTube and other online video-sharing websites have an enormous potential as tool for science, technology and health education and communication and the professional communities from these areas will do well to engage with these communication channels. The results of this research show that people and groups who oppose mainstream scientific positions already gained a strong foothold on such channels and know very well how to use them to their advantage. More research on how such groups use YouTube and other social media sites is urgently needed in order to counter them successfully [12]. In addition, still very little is known on what kind of scientific content can be found on YouTube and other online-sharing websites, who produces and uploads it, and how various users makes sense of it and perceive various types of content on video websites. So far it has been very difficult to obtain randomized samples of YouTube and other social media content and one possible solution advocated in this contribution that could easily be transferred to further research is to use the free software and open network *Tor* for sampling purposes. Further research is also needed on how various other potentially controversial scientific subjects are depicted on YouTube and other video-sharing websites, such as the, for instance, the theory of evolution or information about vaccination. The academic examination of YouTube and online video-sharing in general has just begun. A solid methodological repertoire to study practices and consequences of online video-sharing empirically is urgently required and conceptual and theoretical work needs to draw on various disciplines and interdisciplinary exchange in order to illuminate this interesting and influential social phenomenon.